# A Power Management and Control System for Portable Ecosystem Monitoring Devices

Marcel Balle, Wenxiu Xu, Kevin FA Darras, and Thomas Cherico Wanger

*Abstract*— Recent advances in Internet of Things (IoT) and Artificial Intelligence (AI) technologies help ecosystem monitoring to shift towards automated monitoring with low power sensors and embedded vision on powerful processing units. Vision-based monitoring devices need an effective power management and control system (PMCS) with system-adapted power input and output capabilities to achieve power-efficient and self-sustainable operation. Here, we present a universal power management solution for automated ecosystem monitoring devices, compatible with commonly used off-the-shelf edge processing units (EPUs). The proposed design is specifically adapted for battery-powered EPU systems by incorporating power-matched energy harvesting (EH), a power switch with low-power sleep mode, and simple system integration in an MCU-less architecture with automated operation. We use a 4-month environmental case study to monitor plant growth under 4mg microplastic (MP) exposure, demonstrating that the setup achieved continuous and sustainable operation. In this plant phenology case study, our power management module is deployed in an embedded vision camera equipped with a 5W solar panel and five various environmental sensors. This work shows the usability of the power management board in environmentally relevant use cases and for tasks in agricultural applications.

## I. INTRODUCTION

BIODIVERSITY decline is caused by climate change, agricultural intensification, invasive species, and pollution amongst other threats, and by directly affecting biodiversity-associated ecosystem services, it threatens the quality of human life [1][1], [2] Effectively responding to this situation requires understanding biodiversity change through high resolution spatial and temporal data across taxonomic groups and ecosystems with state-of-the-art technology [3], [4]. Collecting representative long-term datasets and knowledge sharing require the development of cost-effective, portable, nonintrusive, and scalable ecosystem monitoring methods [4], [5]. Emerging technologies such as the Internet of Things (IoT) and artificial intelligence (AI) show promising potential to elevate ecosystem monitoring towards smart devices with real-time and automated operation [6]. For instance, wireless sensor networks (WSNs) use IoT for environmental monitoring and consist of a group of cost-effective, low-power and small-size connected nodes that acquire environmental parameters and transfer the obtained data between each other or a central station [7]. Environmental monitoring applications with WSNs measure scalar physical phenomena such as temperature or humidity, which can only provide limited information about animal and plant richness and abundance [8]. By contrast, time-lapse cameras and remote camera traps are important tools in research fields such as vegetation phenology [9] and wildlife observation [10], and have seen an increasing demand to gather long-term and global-scale standardized image datasets of vegetation and animal species [11]. WSNs can be equipped with image and audio sensors, also called Multimedia WSN [12] to facilitate multi-source data collection and improve spatial coverage. These types of WSNs have already been used in ecological studies, for instance for wildlife tracking [13], [14]. While visual information provides significant advantages in Multimedia WSNs, image sensors and processors require higher power consumption, data rate and flash memory storage compared to traditional WSNs [15]. Sensor network studies have been focusing on various energy efficiency optimization methods in both hardware and software [16] to improve flexibility and scalability of multimedia WSN systems, but the vast amount of data that needs to be processed remains a challenge in environmental and agricultural field-based research.

With recent advances in AI technologies, deep learning and machine vision algorithms are used in ecological studies to enable time-efficient, reliable, and objective analysis of large datasets generated from vision-based monitoring devices [17]. Specifically, AI-based methods provide solutions to overcome current bottlenecks in data collection and analysis [18] and use wireless data transfer to a centralized processing unit for automatic real-time analysis [19]. However, these methods trade off with high power consumption and, hence, make it impractical for functional deployments where stationary power supply is unavailable. The idea of battery-powered automated monitoring with on-board machine vision, so-called embedded vision, can overcome these issues through local image processing and filtering [20] to only transmit valuable data on emerging low power and high-performance hardware platforms [21]. Research fields like environmental sciences and ecology as well as agricultural applications already benefit from embedded systems for instance for real-

*Corresponding authors: Marcel Balle, Thomas Cherico Wanger*

Marcel Balle, Wenxiu Xu, and Thomas Cherico Wanger are with the School of Engineering, Westlake University, Hangzhou 31000, China (e-mail: marcelballe@westlake.edu.cn; xuwenxiu13@westlake.edu.cn; tomcwanger@westlake.edu.cn).

Kevin FA Darras was with the School of Engineering, Westlake University, Hangzhou 31000, China. He is now with the ECODIV Department, INRAE, 45290 Nogent-sur-Vernisson, France (e-mail: kevin.darras@inrae.fr).

2time plant growth detection [22], estimating trends in moth abundance [23], and monitoring ecosystem services such as biological pest control and pollination [24], [25], [26]. Mostly, off-the-shelf edge processing units (EPUs) are used with basic circuit architectures that lack advanced power management such as integrated multi-source battery charging and low-power sleep modes at sub-milliwatt levels, paired with a bulky and inflexible design [27].

Here, we present a universal power management and control system (PMCS) suitable for all single-board computers or EPUs commonly used in ecosystem monitoring studies, focusing on low power, multi-functionality, and practicality for deployments in the field without stationary power supply. The design incorporates only off-the-shelf integrated circuits (ICs) combined with switching circuits in an MCU-less architecture for low cost, automated, and highly flexible operation to provide regulated power output, energy harvesting (EH), and low-power sleep mode. The proposed PMCS architecture provides the following advantages.

1) Solar EH power matched with EPUs.
2) Universal serial bus (USB) connection with On-The-Go communication and fast battery charging.
3) Automatic power source switches.
4) Continuous operation through load sharing circuits.
5) Always-on low-power real-time clock (RTC).
6) Low-power sleep mode with software power-off and scheduled power-on.
7) Regulated 5V power output with high current capability.
8) External devices integration with power supply control.
9) Processor-friendly battery voltage monitoring.
10) Extended protection on USB, solar, battery and output power lines.
11) Panel-mount and waterproof user interface combination.
12) Cost effective and compact format.

In Section II, we compare related studies. Section III then describes requirements and solutions of the proposed PMCS circuits design. In Section IV, we present an environmental science monitoring use case (i.e., monitoring plant growth with microplastic (MP) treatments over a 4-month period) to present the advantages of the PMCS. Finally, in section V we assess the overall performance of the system and conclude with future applications. The abbreviations used in this paper are listed in Table I.

## II. RELATED STUDIES

While many EH and power management solutions exist for self-sustainable WSNs and ecosystem monitoring devices in the low power range [34], only a few studies have integrated these solutions with embedded vision devices with high power requirements. The main features of the related works are presented in Table II. In summary, Homan et al. [28] introduced an insect monitoring trap with high-quality images and various environmental sensors, operating on the Jetson Nano single-board computer by NVIDIA. As the sleep mode power consumption of this board is relatively high at 200mW, the hardware setup included a large 12V, 100W solar panel combined with a commercial charge controller to keep a 100Ah lead-acid battery charged, allowing continuous operation over a full season. Garcia et al. [29] also included such a solar setup in their multimedia WSN design and added a low power microcontroller unit (MCU) to dynamically manage the power supply for the EPU and additional sensors, effectively decreasing the sleep power consumption to 7.3mW. However, the inaccuracy of the MCU internal clock is not practical for long-term operations. To overcome this issue, a dedicated external RTC chip can be implemented for timekeeping and could also allow scheduled power switching as done by Brunelli et al. [30]. Their smart camera setup utilizes the GAP8 system-on-chip [35], achieving 30μW in sleep mode but with low image resolution (320×240, grayscale) and unconventional machine vision models as a trade-off. Mayer et al. [31] designed a comprehensive power management system with multi-source EH and always-on domain through a discrete RTC chip, bringing the sleep power down to 200nW. As this smart power unit is intended for low-power IoT devices, the regulated 3.3V power output for the main system is limited to 300mA, which is not adapted for commonly used EPUs such as the Raspberry Pi or Jetson Nano. Segalla et al. [32] combined a Raspberry Pi with the PiJuice power manager and a 0.8W solar panel, creating an automated pest detection prototype that is assumed to accomplish self-sustainability. Nevertheless, their solution is limited to Pi products which are costly and impractical for portable devices due to their size. Similarly, Suto et al. [33] chose the Raspberry Pi Zero W as the EPU of their insect monitoring prototype and developed a plug-in board to overcome its low power limitations. Along with two communication modules, the proposed design also includes a solar power-tracking charger circuit, three different power rails and an independent RTC circuit, allowing a 22mW sleep mode without the need for an additional MCU. The major drawbacks of the plug-in board are its rather large size, high cost, and, same as some of the previously described automated monitoring prototypes, its power management layer is specifically designed to fit with a designated EPU, making it inflexible and difficult to adapt for other systems and monitoring setups.

TABLE I
OVERVIEW OF THE USED ABBREVIATIONS

| Abbreviation | Description |
|---|---|
| IoT | Internet of Things |
| AI | Artificial intelligence |
| EPU | Edge processing device |
| EH | Energy harvesting |
| MCU | Microcontroller unit |
| WSN | Wireless sensor network |
| PMCS | Power management and control system |
| IC | Integrated circuit |
| USB | Universal serial bus |
| RTC | Real-time clock |
| MP | Microplastic |
| PCB | Printed circuit board |
| VBAT | Battery voltage |
| DC | Direct current |
| PV | Photovoltaic |
| MPPT | Maximum power point tracking |
| FOCV | Fraction open-circuit voltage |
| CCCV | Constant-current constant-voltage |
| $I^2C$ | Inter-integrated circuit |
| GPIO | General-purpose input/output |
| $I^2S$ | Inter-IC sound |
| SPI | Serial peripheral interface |
| ADC | Analog-to-digital converter |
| LED | Light-emitting diode |
| T&H | Temperature & humidity |
| $CO_2$ | Carbon dioxide |

TABLE II
COMPARISON OF AVAILABLE POWER MANAGEMENT SETUPS FOR EMBEDDED SYSTEMS IN AUTOMATED MONITORING

| Paper | VLSID'23 [28] | Sensors'19 [29] | SENSORS'20 [30] | TPEL'21 [31] | MetroAgriFor'20 [32] | Agricultue'22 [33] | This Work |
|---|---|---|---|---|---|---|---|
| Hardware | edge camera | edge WSN | edge camera | PMCS | edge camera | edge camera | PMCS |
| Processor | Jetson Nano | RasPi3 | GAPUINO | sub-W systems | RasPi | RasPi Zero W | universal |
| PMCS size | none | ≈ 76×63mm | 180×100×50mm | 12×12mm | 65×56×20mm | ≈ 90×100mm | 65×30mm |
| PMCS cost | - | ≈ 28 USD | N/A | 11 USD | 57 USD | ≈ 39 USD | 10 USD |
| Power output | - | 3.3V×0.5A 5V×3A | 3.3V×0.3A 5V×1A | 3.3V×0.3A | 5V×2.5A 3.3V×0.1A | 3.3V×1A 5V×3A | 5V×2.4A |
| Sleep power | 200mW | 7.3mW | 30μW | 200nW (54nA) | 2.1mW (PiJuice) | 22mW | 780μW (211μA) |
| Input voltage | - | 7–18V | USB or 7–15V | 3.4–5.5V | 2.5–4.2V | 7.3V | 2.8–4.2V |
| RTC | - | STM32L162 | time & sleep | time & sleep | STM32F030 | time & sleep | time & sleep |
| Configuration | - | MCU software | MCU software | MCU software | MCU software | hardware-set | hardware-set |
| Power switch | - | reset | reset | N/A | 3×power & functions | none | power & functions |
| VBAT read | - | none | N/A | N/A | fuel gauge | none | enable on ADC |
| Charging | solar | solar | yes | nano power | solar & USB/DC | solar | solar & USB |
| Component | PV regulator | PV regulator | N/A | BQ25570 | BQ24160 | LT3652 | CN3791 & IP2312U |
| Power input | 100W | N/A | N/A | 200mW (5V×40mA) | 0.8W 12.5W (5V×2.5A) | 37.78W (17.17V×2.2A) | 50W (18V×2.77A) 15W (5V×3A) |
| Peak efficiency | N/A | N/A | N/A | 93% | 93% | 88% | 94% & 94% |
| Charge current | N/A | N/A | N/A | 285mA | 2.5A | 1.2A | 3A & 2.65A |
| Topology | N/A | N/A | N/A | boost | switch-mode | switch-mode | switch-mode |
| EH MPPT | N/A | N/A | N/A | FOCV | PMW | CCCV | CCCV |
| Mounting | - | screw holes | screw holes | solder pads | screw holes | screw holes | panel connectors |
| Connections | - | I²C, GPIOs, I²S, ADC | ANT-SMA, N/A | I²C, GPIOs, VBAT, N/A | I²C, GPIOs, VBAT, USB | VBAT, ANT-SMA | I²C, 7×GPIOs, ADC, VBAT, USB |

Compared to the hardware configurations from the related studies described above, the proposed PMCS is a standalone design accepting both solar EH and USB power for battery recharging and hosting an always-on RTC chip. Although the 780μW power consumption in sleep mode is not the lowest in this list, it can already significantly reduce the average power consumption of the supplied monitoring system and safe operation is always assured, even when the battery output drops as low as 2.8V. The 5V regulated output feeding up to 2.4A of current is suitable for running computationally heavy algorithms on the main processor. In addition, to further improve flexibility of different systems and simple integration, the layout of the printed circuit board (PCB) accommodates all the necessary input, output, and signal connectors. The PCB also provides a complete external user interface, a multi-function power switch button with panel-mount waterproof USB and multi-pin connectors on the bottom side of the module. All ICs and passive components are placed on the top layer to facilitate mounting, debugging and reduce assembly cost as shown in Fig. 1. Our module size of 65mm by 30mm is compact enough for a plethora of monitoring systems and comes at a low cost of 10 USD.

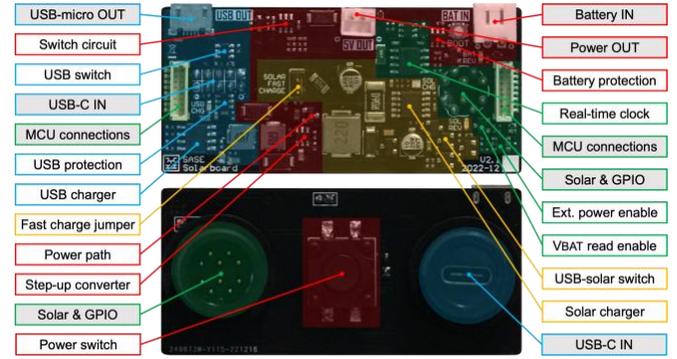

Fig. 1. Top and bottom sides of the assembled PCB module with circuit and connector descriptions.

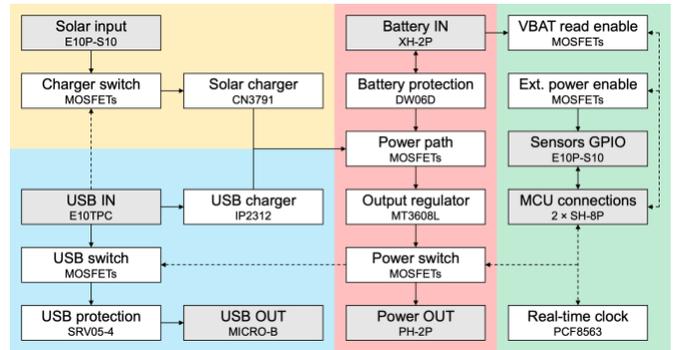

Fig. 2. Simplified block diagram of the PMCS architecture.

III. SYSTEM ARCHITECTURE

Fig. 2 illustrates the PMCS architecture in a simplified block diagram, separated into four parts: solar EH (yellow), USB data transfer and charging (blue), power regulation path (red), and control management (green). White blocks represent IC components or circuits while gray blocks are connectors. Power rails and signal lines are depicted by solid and dashed arrows respectively. The four parts are described in the following subsections along with the connectors and mounting methods.

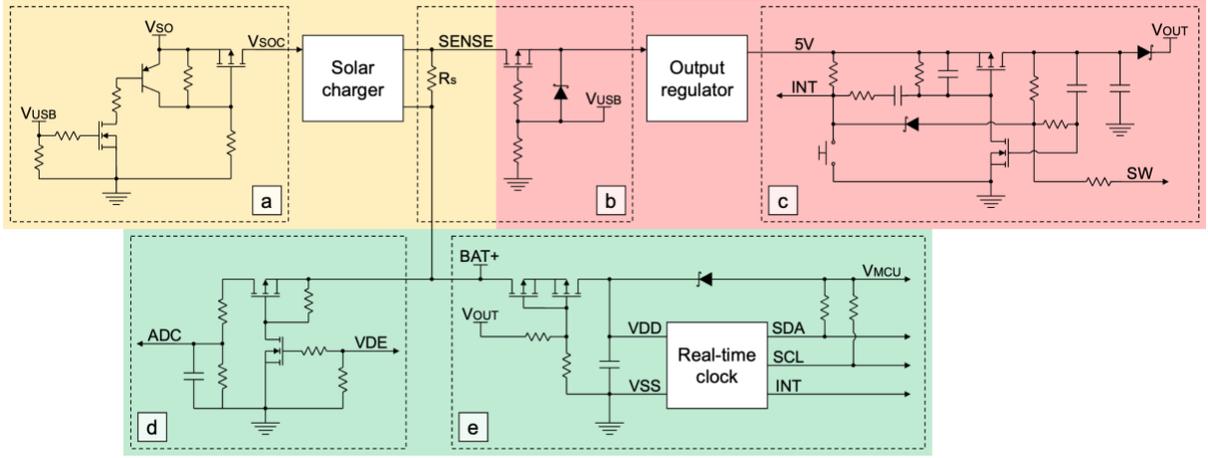

Fig. 3. Schematics of the automatic switching circuits in the PMCS architecture. (a) Solar-USB input auto switch. (b) Power path. (c) Soft latch circuit. (d) ADC-signal enable switch. (e) RTC power supply switch.

*A. Solar Energy Harvesting*

Harvesting energy from natural sources in the environment is an essential function to overcome the first issue of finite energy storage and effectively provide continuous power supply for off-grid devices. The crucial requirement for EH is to match the power input and charge current with the average and peak power consumption of the supplied system. In solar EH, the best efficiency in the energy conversion process is achieved by applying maximum power point tracking (MPPT) on the varying output of the solar panel. To reach self-sustainability, the solar panel power and battery charging capabilities must be large enough to overcome limited hours of sunlight per day, energy conversion efficiency and losses. On the other hand, for low-power applications, necessitating a compact form-factor, the PMCS must also be able to accept energy from smaller solar panels.

According to these requirements, the EH circuit is based on the CN3791 solar power lithium-ion battery charger IC chip utilizing pulse-width modulated switch-mode topology and a current sense resistor to determine the battery charging current. The constant voltage method, used to track the maximum power point of the photovoltaic cell, is set with a resistor divider to accept standard 18-V solar panels as they are available in a wide range of different power outputs and sizes. To make full use of the solar energy at hand while assuring safe charging of the battery, the charge current is fixed at 2A but can be elevated to 3A with a jumper on the marked header pins. A bi-color light-emitting diode (LED) indicates the charging status of the chip: red while charging and green when the battery is full. The circuit hosts additional transistors and components to prevent battery current backflow and to create a cut-off switch in case of a reverse polarity connection on the solar power input, triggering a warning LED indicator. When the system is powered through the USB connector on the board, the transistor-based automatic switch, shown in circuit (a) of Fig. 3, disconnects the solar power input from the charger to avoid hazardous charging behaviors when both power sources are present at the same time. This circuit prioritizes power from the USB input over solar, as it is more stable and can provide more current.

*B. USB Data Transfer and Charging*

Including a USB connector interface facilitates access to and retrieval of the collected data. The USB connector also allows changing the EPU program or operation parameters without uninstalling or opening the monitoring device. Moreover, USB input provides a reliable power source that can be drawn to charge the energy storage. As most newly-developed EPU and MCU boards feature direct USB programming and can be powered by a 5V supply, the circuitry simply relays power and universal asynchronous receiver-transmitter data lines from input to output of the board.

The PMCS architecture incorporates the IP2312 synchronous switch step-down charger IC, to enable charging the battery through a USB type-C connector. The maximum charge current in constant current mode is set to 2.65A by an external resistor while the trickle charge current is fixed at 100mA. During the charging process, a blue LED blinks at 0.5Hz and switches to constant light when the battery is fully charged.

The remaining blocks in this part are components for current surge protection on the data lines and a transistor switch, controlled by the main 5V output to power-on the USB-micro output when the push button is pressed. Compatibility between type-C input and micro-B output is achieved by pulling the CC pin to ground with a 5.1kΩ resistor. This allows data transfers with the processor through the USB connector on the PMCS.

*C. Power Regulation Path*

**Battery Protection:** First, to ensure safe voltage and current flow from and to the battery while charging or discharging it, the DW06D lithium-ion polymer battery protection chip provides the fundamental over-charge, over-discharge, over-current and short circuit protections. A tiny push button is incorporated to reset the battery protection chip when battery power is connected for the first time and the discharge function is disabled. The battery input connector is accompanied by electrostatic discharge protection diodes and a reverse polarity LED indicator with cut-off switch.

**Load Sharing:** Falsified charging current behaviors caused by a direct connection between charger output, battery and system load should be avoided by implementing a load sharing topology. In the PMCS design, a hybrid load sharing circuit is created to continuously supply the output load with the required current regardless of solar or USB power presence and to maintain normal battery charging cycles. For the USB charger, a standard transistor-based power-path selection circuit will directly power the system load from USB power when it is present [36]. As for the solar charger, an IC chip with sense resistor-based charge current is used to connect the load before the sense resistor [37]. Consequently, the charging current and current sense are not affected by the system load so that any current can be drawn at any time, irrespective of solar input power. By combining these two power path topologies, as shown in circuit (b) of Fig. 3, the power sources can simultaneously provide the output load and charge the battery at the same time if the load current draw is lower than the maximum set charge current. In the case where load current is higher than charging current, both the power input source and the battery will provide power for the output load.

**Output Regulator:** Because of the chemistry of lithium-ion batteries, their voltage can range from 4.2V down to the over-discharge voltage threshold, which is 2.4V here. To ensure proper operation of all the components and circuits, the battery voltage is stepped up to 5V by the current mode boost converter MT3608L with a current protection limit set to 2.4A. With a high efficiency up to 98% at low load currents and 1.2MHz switching frequency, the regulator is fitting for low power modes such as the sleep mode, and for a wide range of power loads. The regulator also maintains a stable output voltage when switching between high and low current draws. Other advantages of this chip are its 2.2V minimum operation voltage so that it can provide the desired 5V output during the entire battery life, and a typical quiescent current of 200μA which is low enough to allow the regulator to keep running during sleep mode.

**Power Switch:** Many EPUs lack the integration of a power switch to turn the power supply for the main system on and off, which is useful for saving energy and avoiding unstable power cuts whenever the power connector is manually removed. As a result, the PMCS architecture includes a push button combined with a soft latching circuit allowing to turn on the power output with a single button press and turn it off with a long press [38]. At the same time, the button press can be detected by the main controller unit on a signal line and power can be turned off by pulling this same line low for a few seconds. With this feature, it is possible to program safe self-shutdown of the main system after the completion of critical tasks or at a certain time, and a vast variety of software-coded functions on different button press combinations such as single press or double press. The power switch circuit (c) in Fig. 3 is composed of metal-oxide-semiconductor field-effect transistors, whose behavior is defined by their gate-source voltage, and passive components with values adapted to a fixed voltage. The circuit is placed after the output regulator so it can function with the stable 5V rail, ensuring flawless operation irrespective of battery voltage level.

*D. Control Management*

**Battery Voltage Reading:** Monitoring the battery voltage gives the possibility for the processor to automatically manage and optimize EH, and power consumption through operation adaptation. For example, when the battery voltage falls below a critical threshold, the device can be programmed to send a warning to the user before reducing its duty cycle or going into sleep mode. As the maximum battery voltage of 4.2V is too high for most analog-to-digital converter (ADC) inputs, and processors often have a limited amount of these peripherals, the circuit (d) shown in Fig. 3 comprises a resistor-based voltage divider enabled by a signal line. In this way, the divider is turned off when not used, which reduces power consumption and allows the ADC port to be used for other measurements or alternative functions.

**External Devices Supply:** The power management system can provide power for the main system, but also improve external hardware through effective transfer of various communication protocols on multiple signal lines from the processor to external modules. The PMCS includes an enable switch on the ground rail for external circuits such as sensors, lights, servo motors or relays such that the processor can turn on their power supply only when needed, allowing further optimization of the overall power consumption.

**Real-time Clock:** Environmental science applications often rely on long-term sampling, for instance to monitor plant growth (i.e., phenology monitoring applications; see case study in Section IV). These phenology cases are intended to operate at least for several months ideally without interruption to change batteries or reset software parameters. It is, hence, critical for any long-term application to have an accurate time clock, which is not the case for many processors; for instance, MCUs are running on low clock frequencies or entirely shutting down their internal RTC peripheral when entering a low-power sleep mode, which means low accuracy or complete loss of the current time. An always-on, low power external RTC chip can provide uninterrupted and accurate timing, suitable for such applications. Some of these chips feature an interrupt pin that is triggered after or at a specific pre-set time, suitable to wake up processors from sleep modes. However, with the lack of sub-mW low power modes on EPUs or MCU modules capable of embedded vision, the most beneficial way of utilizing the RTC interrupt output is combining it with a switch circuit to completely cut off the power supply to the main system and turn it back on automatically through the triggered signal. In this case, the power consumption is lowered to the quiescent current of the few always-on components in the power management system.

The PCF8563 RTC chip was chosen because it is commonly used in IoT devices, making it low-cost and simple to program through the many existing libraries. Optimized for low power consumption, the chip has a typical 400nA quiescent current, suitable for this design. The RTC is powered by the same supply voltage as the processor in the main system to ensure stable Inter-integrated circuit (I²C) communication between both chips. As secondary power supply, the main lithium-ion batteries are used to avoid an additional spacious coin cell battery on the board layout. When available, the primary power supply must be prioritized over the secondary one to allow proper reading and writing operations with the main processor. This function, along with the reverse voltage flow protection, is achieved by switching a double P-channel transistor with the generated 5V output as illustrated in circuit (e) of Fig. 3. In this way, the power supply switch-over is always successful, regardless of battery voltage level.

The hardware setup to initialize the scheduled power-on is established by directly connecting the interrupt pin of the RTC chip to the pulled-up side of the push button in the soft latching circuit. As for the software side, the interrupt output of the PCF8563 generates a low signal every programmed countdown period and, when set as a pulsed signal, it acts the same as a short button press, effectively latching on the circuit and powering the output. During this sleep mode, only the voltage regulator, the protection circuits and the RTC along with some transistor and resistor components consume power which results in a current consumption of about 211µA.

*E. Connections*

**Internal:** Battery input and 5V power output are provided through a unidirectional XH connector with 2.54mm pitch and a PH-2.00mm connector, respectively. Utilizing two different connectors is an additional protection that prevents potential reverse and mixed-up connections. Power rails and general-purpose input-output signals used to communicate with the RTC, detect a button press and execute a self-shutdown, measure battery voltage, enable and communicate with external devices, are transferred through two surface mount 8-pin SH-1.0mm connectors. These connections increase controllability and operation flexibility.

**External:** Solar power from the photovoltaic panel can be supplied through a 10-pin connector that also includes two power rails, ground, and five signal lines with I²C and ADC main functions for external circuits and devices. Considering that solar panel and sensors are typically set up outdoors, both the 10-pin and the USB type-C input connectors are waterproof and can be panel-mounted inside a sealed container to effectively separate and protect the electronic parts from the natural elements. These two off-the-shelf connectors are placed on each side of the centered push button on the bottom side of the PCB module. When installed with a sealing rubber cap, the base of the button is pressed evenly against the panel surface, ensuring waterproofing while allowing access to the power switch functions from outside. An example of the PMCS mounted inside a housing structure is shown in Fig. 5(b).

IV. EXPERIMENTAL RESULTS

We investigated the long-term functionality of the designed PMCS in a practical use case where we monitored plant growth of classical horticulture Dianthus flowers (*Dianthus*

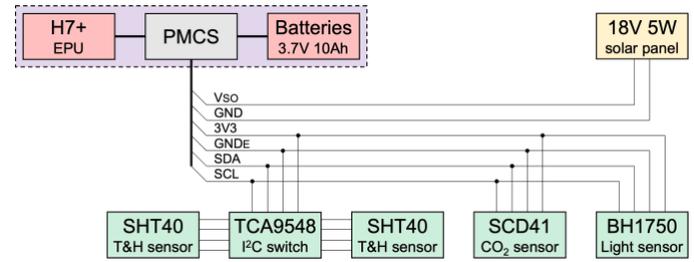

Fig. 4. Block diagram of the electrical elements and their wiring in the monitoring setup.

*carthusianorum*) under the effect of plastic mulching over a period of 4 months. This use case represents a common phenological monitoring case that can be used in agricultural studies. For this, we integrated our PMCS in an embedded vision camera equipped with the openMV H7 Plus board in a waterproof housing [26]. The system shown in Fig. 5(b), is powered by 3 internal parallel lithium-ion batteries with a total capacity of 10Ah and recharged with an external 270mm by 180mm solar panel harvesting up to 5W.

The embedded vision camera captured top-view images of the plants every 30 minutes from sunrise to sunset and computed their abundance through a simple blob-based color-tracking algorithm. Moreover, three additional sensors measured temperature and humidity (Sensirion SHT40) above and 50mm underneath the soil surface, carbon dioxide ($CO_2$) concentration (Sensirion SCD41) and ambient light (Rohm Semiconductor BH1750). These sensors communicate data through the I²C protocol, allowing multiple unique devices on the same bus. For the two temperature and humidity sensors, an I²C switch (Texas Instruments TCA9548) is implemented to read their measurements separately. Communication signals and supply rails for all the sensors, and power from the solar panel are fed to the camera through the 10-pin waterproof connector as shown in Fig. 4.

Plastic mulching has been shown to increase soil temperature and humidity, and to result in better yield [39]. It is predicted to change ecosystem services across multiple spatial scales [40]. Six plant monitoring boxes were constructed and then randomly allocated into control (i.e., untreated soil; n=3) and MP treatments (i.e., soil spiked with MP concentration of 4g/l soil; n=3). The plastic particles size of up to 0.5mm was chosen based on existing studies of known particles sizes to affect plant growth [41]. Before mixing in plastic particles, the total amount of soil was mixed and then distributed equally across all six boxes to minimize variation from different lighting and humidity conditions. The boxes were placed 20cm apart from each other to assure statistical independence of boxes as shown in Fig. 5(a). The experiment took place in Hangzhou, China (30.34257°E, 120.03787°N), from September 2023 to January 2024 covering the summer, autumn, and winter in Hangzhou. In case of insufficient rain, we watered the boxes every 3 days.

With this setup, environmental data can be collected with high temporal resolution and accuracy. For each box, the recorded lithium-ion battery voltage shown in Fig. 6(a) never falls below 4V, showing that the system achieves self-sustainability, allowing for continuous data collection. It should be noted that due to a bad connection with the sensors, box1 did not record all the data from October 4th until 10th, and box4 from October 23rd until November 3rd. As expected, the measurements of the ambient humidity, luminosity, and temperature in Fig. 6(b), (c), (d) show similar values for all boxes. Air and soil temperature, illustrated in Fig. 6(d), shows the foreseen higher values in the air and no significant difference between control and MP treatment. In Fig. 6(e), $CO_2$ measured above the plants show no significant differences but lower values in untreated soil (blue line) compared to the MP treatment (red line). Relative plant growth extracted from the computed green pixel count in Fig. 6(f), is significantly faster in the control compared to the treatment boxes.

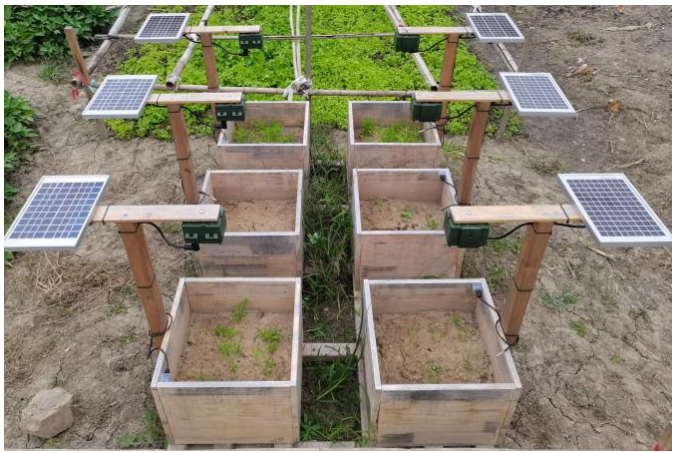

(a)

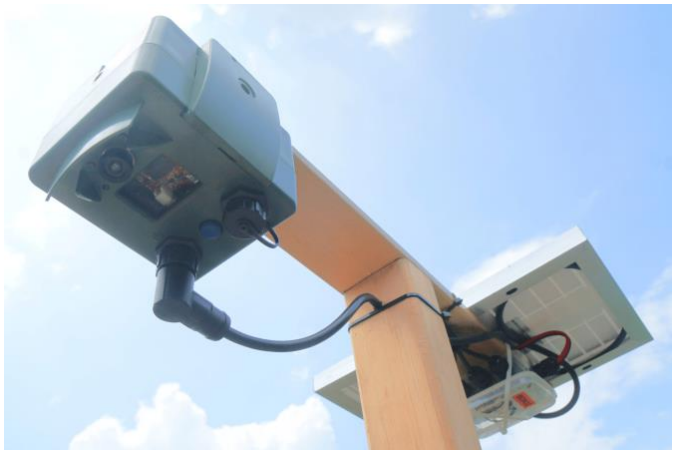

(b)

Fig. 5. MP effects on plant growth - case study. Monitoring boxes. (a) setup in the field. (b) embedded vision device installation.

## V. CONCLUSION

This paper presented the design of a versatile PMCS for portable embedded vision-based monitoring systems that have recently gained traction in ecosystem monitoring applications. Based on the requirements for each section in the architecture, the components and wiring are described, accompanied by schematic circuit drawings. By adopting power characteristics matched with the demand from the processing system and incorporating efficient power-saving features, most notably the 211µA sleep mode, continuous supply and self-sustainability could be achieved. The inclusion of waterproof panel mount connectors provides a simple integration method and a complete user interface. The main advantages of this design, compared to the existing power management solutions for ecosystem monitoring EPUs, are the multi-source battery charging, simple integration for different systems, flexibility to be adapted for various applications, and cost.

Our experimental case study shows how the PMCS integrated in an embedded vision device combined with solar energy harvesting and several sensors can enable comprehensive environmental data collection with one device. In our case study, we could show the effects of plastic pollution under natural concentrations on plant growth and soil fertility of an important horticultural plant. More generally, it verified the use of the PMCS in long term monitoring applications in environmental science in natural and agricultural systems. In conclusion, our PMCS may play an important role in the advancement of embedded vision devices in ecosystem monitoring applications.

### DATA AVAILABILITY

The data generated from the experimental case study of this work is openly available at http://ieee-dataport.org/12758.

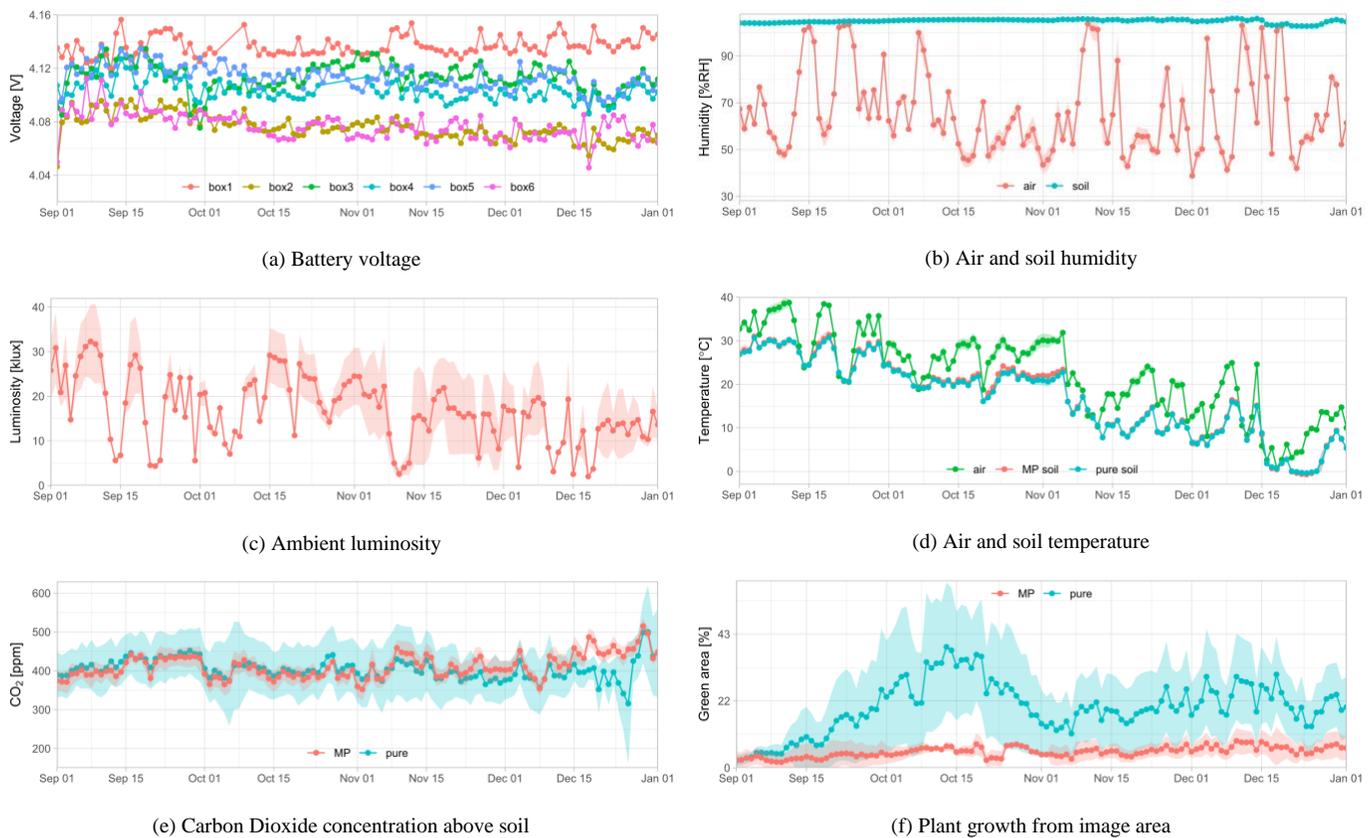

Fig. 6. Graphs of the daily average values with standard deviation from the sensor measurements and analyzed images taken between 7:00 and 16:30 over all boxes and the two soil treatments.

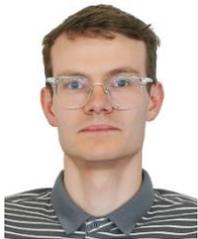

**Marcel Balle** received the B.S. degree in systems engineering from the HES-SO Valais-Wallis University of Applied Sciences, Sion, Switzerland, in 2018, and the M.S. degree in electronics science and technology from Zhejiang University, Hangzhou, China, in 2022.

He is currently a Research Assistant with the Sustainable Agricultural Systems and Engineering Laboratory, Westlake University, Hangzhou, China. His research interests include vision-based environmental monitoring, embedded and intelligent systems integration for ecological applications, low power sensor and microcontroller circuit design.

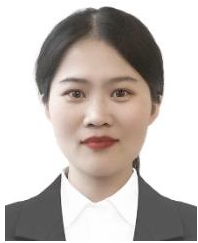

**Wenxiu Xu** received the B.S degree in forestry from Huazhong Agriculture University, Wuhan, China, in 2014, and the M.S. degree in ecology from the University of Chinese Academy of Sciences, Beijing, China, in 2017. She is currently working towards the Ph.D. degree in environment science and technology with Westlake University, Hangzhou, China.

Her research interests include sustainable food system transformation, ecological technology development and application.

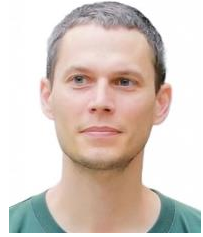

**Kevin FA Darras** received the B.S. degree in biological sciences from the Swiss Federal Institute of Technology, Zürich, Switzerland, in 2007, the M.S. degree in ecology, biodiversity, and evolution from the University Pierre et Marie Curie, Paris, France, in 2010, and the Ph.D. degree in agricultural sciences from the University of Göttingen, Göttingen, Germany, in 2016.

From 2020 to 2022, he was a Postdoctoral Researcher with the Sustainable Agricultural Systems and Engineering Laboratory, Westlake University, Hangzhou, China. He is currently a permanent Researcher with the research group EFNO, INRAE, France. His research interests include agroecology, community ecology, biodiversity monitoring, ecoacoustics and embedded machine vision.

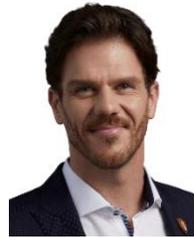

**Thomas Cherico Wanger** received the B.S. degree in biology and M.S. degree in ecosystem management from the University of Tübingen, Tübingen, Germany, in 2005 and 2007, respectively, and the Ph.D. degree in environmental sciences from the University of Adelaide, Adelaide, Australia, in 2011.

From 2011 to 2018, he did Postdocs with Stanford University, Stanford, CA, USA, the Swedish Agricultural University, Uppsala, Sweden, and the University of Göttingen, Göttingen, Germany. He has received multiple awards for scientific excellence and his work has been cited over 10,500 times. Besides his research career, he held senior management positions in the private sector and founded ecoNect Ltd, a Hangzhou-based company that brings applied AI tools to environmental research and education. He is also a regular keynote speaker at high level conferences and is available as a consultant for topics around sustainable agriculture, agroforestry, and AI.

Since 2019, He has been an Associate Professor with the School of Engineering, Westlake University, Hangzhou, China, where he leads the Sustainable Agricultural Systems and Engineering Laboratory. His main research interest is on diversified agricultural systems design for sustainable and profitable agriculture globally. His Team uses AI, embedded systems, and spatial modelling approaches to support a global food systems transformation in research and education. He has established two large-scale research platforms including the Global Agroforestry Network for cocoa, coffee, and rubber research and the China Rice Network for diversified rice production in China.